\documentstyle[preprint,aps]{revtex}
\tighten
\begin{document}
\draft
\preprint{\hfill NCL96--TP2}
\title{BRST Invariant Boundary Conditions for Gauge Theories}
\author{Ian G. Moss and Pedro J.Silva}
\address{
Department of Physics, University of Newcastle Upon Tyne, NE1 7RU U.K.
}
\date{March 1996}
\maketitle
\begin{abstract}
A systematic way of generating sets of local boundary conditions on the
gauge fields in a path integral is presented. These boundary conditions
are suitable for one--loop effective action calculations on manifolds
with boundary and for quantum cosmology. For linearised gravity, the
general proceedure described here leads to new sets of boundary
conditions.
\end{abstract}
\narrowtext
\section{INTRODUCTION}

The aim of the work reported here is to characterise sets of local
boundary conditions on the fields in a path integral. This is a
non-trivial problem for gauge theories, where the boundary conditions
have to be consistent with the gauge symmetry. In the BRST approach
\cite{brs,ty}, which we examine, this consistency with the gauge
symmetry translates into BRST invariance. The gauge fields are
augmented by extra families of ghosts, antighosts and auxillary fields
that also require boundary conditions.

Boundary conditions are needed for effective action calculations on
manifolds with boundary and for the evaluation of wavefunctions in
quantum cosmology \cite{hawking,halliwell,luoko}. In many of these
applications the geometry is curved, and this is where local boundary
conditions are especially useful. Boundary conditions that required
separating transverse from longitudinal photons, for example, would be
non-local. This would not be a problem in flat spacetime, because the
separation is local in momentum space. In curved spacetimes, however,
these non-local operations are best avoided.

There is another important reason for considering local boundary
conditions. To first order in Planck's constant, the result of a path
integral is closely related to the asymptotic behaviour of the
eigenvalues of an operator. With local boundary conditions, the
asymptotic behaviour of the eigenvalues is determined by local tensors
through a heat kernel expansion \cite{g1,luckock,moss1,moss2,g2,v1}.

Local boundary conditions have been described before for Maxwell gauge
theory, where the fields of interest include perturbations in the
vector potential, ghosts and antighost fields \cite{moss3}. There are
two sets of boundary conditions corresponding to fixing the magnetic or
electric field on the boundary. Each set has mixtures of Dirichlet and
Robin boundary conditons. If we split the fields into two subsets by
using projection operators $P_\pm$,
\begin{eqnarray}
\left({\cal L}+\psi\right)\,P_+\phi&=&0\label{rn}\\
P_-\phi&=&0,\label{dt}
\end{eqnarray}
where ${\cal L}$ is the Lie derivative along the normal to the boundary
$\Sigma$ and $\psi$ is a matrix. These boundary conditions are now
widely used \cite{moss4,poletti,e1,e2,b1,b2,e3,k1,k2,k3}.

A similar set of boundary conditions was found for gravitational fields
\cite{moss4,l2}, but it was soon discovered that this set of boundary
conditions was not invariant under BRST transformations
\cite{moss3,e4}. A set found by Barvinski \cite{b3} is invariant under
BRST, but not quite of the same form. In this set, $\psi$ in equation
(\ref{rn}) includes a first order differential operator restricted to
the boundary \cite{e4,e5,e6,e7}.

The gauge-fixing in both of the cases mentioned above is a covariant
function of the gravitational background. By contrast, allowing
non-covariant gauge-fixing allows a set of boundary conditions that is
both BRST invariant and of the mixed type \cite{e7}. These
non-covariant approaches are not applicable, so far, to all topological
situations. Other possibilities have also been considered \cite{e8,mar}

It appears that gravity with covariant gauge-fixing terms in the
Lagrangian requires us to generalise the original class of mixed
boundary conditions to new classes ${\cal M}_n$, where $\psi$ is a
differential operator of order $n$. The asymptotic behaviour of the
heat kernel is known for mixed boundary conditions ${\cal M}_0$
\cite{moss6}. It should be possible to extend these results to classes
${\cal M}_1$ and ${\cal M}_2$ without too much difficulty.

In the next section we shall see that a set of boundary conditions of
type ${\cal M}_n$ can always be generated, based upon a standard idea
of having the ghost and antighost fields vanish on the boundary
\cite{h1}. We shall also see how this gives rise to a means of
generating new sets of boundary conditions through the application of
canonical transformations between the ghosts and antighosts.

For linearised gravity with t'Hooft--Veltman gauge-fixing (sometimes
called harmonic gauge)\cite{thv}, the general proceedure described
above leads to two new sets of boundary conditions in class ${\cal
M}_2$. With certain restrictions on the extrinsic curvature of the
boundary, one new set of boundary conditions arises that is ${\cal
M}_0$ and is therefore the first BRST invariant set of boundary
conditions of the original mixed type.

In this paper we will the signature of the background four-metric to be
(++++).

\section{VANISHING GHOSTS}

In the BRST approach to the path integral the original fields $q$ are
augmented by ghosts $c$, antighosts $\overline c$ and auxilary fields
$b$ (see \cite{h1} for a revue). The path integral over the fields on a
manifold with boundary $\Sigma$ will result in an amplitude in which
the fields are specified on $\Sigma$,
\begin{equation}
\Psi=\Psi(q,c,\overline c,b\,;\Sigma).
\end{equation}
If $\Sigma$ has only one connected component, then the amplitude would
be a wave-function in the sense adopted in the study of quantum
cosmology \cite{hawking}.

When evaluating the path integral, a classical term is usually
subtracted from the fields so that the residual fields satisfy
simplified boundary conditions. The result of the path integral can
then be written in terms of operators acting on the fields.

Our aim is to find what additional restrictions have to be placed on
the fields in order to recover the correct number of physical degrees
of freedom. In most applications, for example, an immediate restriction
follows from the elimination of the auxilliary field, leading to
boundary conditions $b^i=\overline E^i(q,{\cal L} q)$, where ${\cal L}$
is the Lie derivative along the normal to the boundary.

We will regard events on the boundary as simultaneous and ${\cal L}q$
as the time derivative of $q$. The importance of these time derivatives
indicates that Hamiltonian methods should be useful.

In the classical Hamiltonian approach we introduce the Poisson
brackets,
\begin{eqnarray}
\left[q_n,p^m\right]_{PB}&=&\delta_n^{\ m},\nonumber\\
\left[c_i,p^j\right]_{PB}&=&-\delta_i^{\ j},\nonumber\\
\left[\overline c^i,\overline p_j\right]_{PB}&=&-\delta_i^{\ j}.
\end{eqnarray}
The momenta are distinguished by their indices, $m$ and $n$ for the
fields and $i$ and $j$ for the ghosts. For field theories, the index
also includes the coordinates on $\Sigma$ and summation over a repeated
index includes integration over $\Sigma$.

Two important operators that we shall use are constructed from
classical generating functions \cite{h1}. The ghost-number generator
keeps track of the number of ghosts,
\begin{equation}
G=c_ip^i-\overline c^i\overline p_i.
\end{equation}
The BRST generator $\Omega$ generates BRST symmetries $s$,
\begin{equation}
s^Rz=[z,\Omega]_{PB}
\end{equation}
where $s^R$ is used to denote BRST acting from the right. The BRST
generator depends on constraints $E^i(q,p)$ and their structure
constants $C^{ij}_{\ \ k}$. In the type of theory known as rank 1 the
gauge-fixed action leads to a BRST generator whic has the form
\begin{equation}
\Omega=\overline p_i\overline
E^i+c_iE^i+\case1/2c_ic_jC^{ij}_{\ \ k}p^k.\label{q}
\end{equation}
We shall assume that the theory has rank 1 for notational convenience.

Vanishing ghost-number and BRST invariance are imposed as fundamental
requirements on the quantum theory. In terms of operators and
amplitudes we set
\begin{eqnarray}
G\Psi&=&0,\label{gc}\\
\Omega \Psi&=&0.\label{qc}
\end{eqnarray}
These conditions, which reduce the space of states to those that may be
regarded as physical, serve as boundary conditions on the path
integral.

The simplest way to satisfy the constraints (\ref{gc}) and (\ref{qc})
is to set the ghost fields to zero on the boundary of the path
integral. The Poisson bracket
\begin{equation}
[\overline c^i,Q]_{PB}=-\overline E^i,
\end{equation}
when expressed as a commutator acting on equation (\ref{qc}) implies
that $\overline E^i$ also has to vanish on $\Sigma$. The set of
boundary conditions so far is therefore
\begin{equation}
c_i=\overline c^i=b^i=\overline E^i=0.
\end{equation}

The BRST variation of the fields $q$ when $c=0$ is given by,
\begin{equation}
s^Rq_n=\overline p_j[q_n,\overline E^j]
\end{equation}
Those fields which commute with $\overline E$ belong to a set we call
${\cal Q}$ and can be fixed on the boundary. The boundary conditions on
the fields which do not commute with $\overline E$ are determined by
the vanishing of $\overline E(q,p)$. The vanishing-ghost boundary
conditions on the gauge-fixed path integral are therefore
\begin{eqnarray}
c_i=\overline c^i=0\\
b^i=\overline E^i(q,p)=0\\
q\hbox{ fixed for }q\in{\cal Q},
\end{eqnarray}
where ${\cal Q}$ is the set of fields whose momenta do not appear in
the gauge-fixing functions $\overline E$. These boundary conditions are
invariant under BRST transformations by construction.

Our principal concern is to list all of the possible sets of boundary
conditions, subject to specific restrictions. An obvious place to begin
is the division of phase space into ghosts and their conjugate momenta,
which is not preserved by canonical transformations (defined below).
One way of creating further sets of boundary conditions would therefore
be to perform an arbitrary canonical transformation before applying the
vanishing-ghost conditions.

In actual fact, not all canonical transformations turn out to be
suitable. Some lead to vanishing-ghost boundary conditions that are not
BRST invariant. This is due to structural changes in the BRST generator
$\Omega$. Because of this fact, we consider a restricted class of
transformations that satisfy the following conditions:
\begin{itemize}
\item[1] The transformation is canonical.
\item[2] It preserves the number of ghosts.
\item[3] The variation of a vanishing ghost vanishes.
\end{itemize}

Condition (3) means that $[c,\Omega]=0$ when $c=0$, where $c$ is the
new ghost field. This condition arises from requiring that setting the
BRST variation of $c$ to zero should not imply any further restrictions
on the fields.

We will consider how these restrictions apply to transformations
between the ghosts, antighosts and auxiliary fields. For this purpose
it is convenient to blur the distinction between ghosts and antighosts
and write,
\begin{equation}
\eta_i=\pmatrix{c_i\cr\overline p_i}\quad\quad
{\cal P}^i=\pmatrix{p^i\cr\overline c^i}\quad\quad
{\cal E}^i=\pmatrix{E^i\cr\overline E^i}.
\end{equation}
We also define $\lambda_i=\lambda_i(q,p)$ to be the set of fields
canonically conjugate to $b^i=\overline E^i(q,p)$.

Canonical transformations from $\{\eta,{\cal P},\lambda,b\}$ to
$\{\eta',{\cal P}',\lambda',b'\}$ are generated by $F(\eta',{\cal
P},\lambda',b)$,
\begin{eqnarray}
\eta_i={\partial F\over\partial {\cal P}^i}&\quad&
\lambda_i=-{\partial F\over\partial b^i}\nonumber\\
{\cal P}^{\prime i}=-{\partial F\over \partial\eta'_i}&\quad&
b^{\prime i}=-{\partial F\over \partial\lambda'_i}.\label{dif}
\end{eqnarray}

The ghost-number operator can now be written,
\begin{equation}
G=\eta_i{\cal P}^i={\partial F\over\partial {\cal P}^i}{\cal P}^i.
\end{equation}
In the new coordinate system,
\begin{equation}
G'=\eta'_i{\cal P}^{\prime i}=-\eta'_i{\partial F\over
\partial\eta'_i}.
\end{equation}
Setting $G'=G$ therefore leads to transformations of the form
\begin{equation}
F\equiv F({\cal P}^i\eta'_j,\lambda',b).\label{f}
\end{equation}

The allowed linear transformations on the ghost and antighost fields
are covered by the following theorem:

Transformations generated by
\begin{equation}
F={\cal A}_i^{\ j}(\lambda'){\cal P}^i\eta'_j+b^i\lambda'_i,
\end{equation}
where the matrix ${\cal A}$ has the following properties
\begin{equation}
{\cal A}=\pmatrix{A&B\cr C&D},\quad dA=dC=dD=0,\quad
dB_i^{\ j}=\case1/2C^{kl}_{\ \ i}B_l^{\ j}\,d\lambda'_k,\label{be}
\end{equation}
satisfy conditions 1--3.

These transformations are manifestly of the form given in equation
(\ref{f}). The rest of the proof is by direct application of eqs.
(\ref{dif}). These allow the generator $\Omega$ to be written in the
form
\begin{equation}
\Omega={\cal A}_i^{\ l}\eta'_{\,l}{\cal E}^i+
\case1/2{\cal A}_i^{\ l}{\cal
A}_j^{\ m}\eta'_{\,l}\eta'_{\,m}C^{ij}_{\ \ k}{\cal P}^k.
\end{equation}
We are also able to replace $b^i$ by $b^{\prime i}$,
\begin{equation}
{\cal E}^i={\cal E}^{\prime i}+
{\partial B_j^{\ k}\over\partial\lambda'_i}p^j\overline p'_k.
\end{equation}
The linear term in $\Omega$ commutes with $c'_i$ and gives no further
boundary conditions. The condition on $B$ in equation (\ref{be})
removes terms beginning $\overline p_i\overline p_jC^{ij}_{\ \ k}$,
which are the only ones that violate condition (iii). This completes
the proof of the theorem.

We can now write the vanishing-ghost boundary condition in terms of the
original variables and obtain new sets of boundary conditions,
\begin{eqnarray}
B_i^{\ j}p^i+D_i^{\ j}\overline c^i&=&0\label{bcgen}\\
B_i^{\ j}\overline p^i-D_i^{\ j}c^i&=&0\\
B_i^{\ j}E^i+D_i^{\ j}\overline
E^i&=&\case1/2D_i^{\ j}C^{il}_{\ \ k}p^kc_l\\
q\hbox{ fixed for }q&\in&{\cal Q},
\end{eqnarray}
where ${\cal Q}$ is now the set of fields whose momenta do not appear
in the other boundary conditions.

For an abelian theory, these boundary conditions allow any linear
combination of the ghosts and their momenta can be set to zero as long
as it is consistent with ghost-number. The remaining boundary
conditions are then determined uniquely. These boundary conditions
therefore include all the possible sets of linear boundary conditions.
Since quantum field theories are effectively abelian up to order
$\hbar$, these also exhaust sets of linear boundary conditions for
one--loop quantum field theory.

There is still one further restriction to impose, namely that the
boundary conditions are in a class ${\cal M}_n$ of the mixed boundary
conditions mentioned in the introduction. This means that the linear
transformations must now be local, and the momenta should be replaced
by normal derivatives. We can then proceed by the following rules:

\begin{itemize}
\item[1] Each set of linear combinations of the constraints $E(q,{\cal
L}q)$ and gauge-fixing conditions $\overline E(q,{\cal L}q)$ that can
be written in the form of equations (\ref{rn}) and (\ref{dt}), possibly
after removing an overall surface-derivative, defines a set of mixed
boundary conditions.
\item[2] The boundary conditions on the ghosts are fixed by equation
(\ref{bcgen}) once the linear combinations are given.
\item[3] The set ${\cal Q}$ of fields that can be fixed on the boundary
are finally identified by examining the combinations $E(q,p)$ and
$\overline E(q,p)$ for any missing momenta.
\end{itemize}

The resulting sets of boundary conditions depend on the choice of
gauge-fixing term in the action. This is to be expected, because the
path integral is usually expressed in terms of operators which
themselves depend on the choice of gauge fixing term. On the other
hand, there is still some freedom in the choice of Lagrangian density
even when the gauge-fixing term is fixed. For example, it is possible
to eliminate the auxilliary field $b$ from the action at an early stage
or to leave it in. This affects the form of the constraints,
gauge-fixing condition and even the momenta, but it does not affect the
final form of the boundary conditions.

\section{Electrodynamics}

A simple example of the preceeding ideas is provided by electrodynamics
in curved spacetime. For Lorentz gauges, the Maxwell field $A_a$ is
accompanied by one ghost field $c$ and one antighost $\overline c$.

In order to set up a phase space associated with the hypersurface
$\Sigma$ we need to decompose the Maxwell field into normal and
tangential components,
\begin{equation}
A_a=\phi_a+\phi\,n_a.
\end{equation}
(The index structure alone distinguishes different vector and scalar
quantities. We find this preferable to a profusion of notation.)
Momenta conjugate to $\phi_a$, $\phi$, $c$ and $\overline c$ are
denoted by $\pi^a$, $\pi$, $p$ and $\overline p$ respectively. The
extrinsic curvature will be denoted by $K_{ab}$ and a vertical bar
denotes covariant differentiation in $\Sigma$.

The Lagrangian density $L$ can be taken to be the sum of three terms,
the Maxwell, ghost and gauge-fixing terms
\begin{eqnarray}
L_A&=&-\case1/4F_{ab}F^{ab}\nonumber\\
L_{gh}&=&-\overline c^{;a}c_{;a}\nonumber\\
L_{gf}&=&-bA_a^{\ ;a}+\case1/2b^2.\label{edl}
\end{eqnarray}
The field $b$ can be eliminated by
\begin{equation}
b=A_a^{\ ;a},\label{aux}
\end{equation}
restricting the nilpotency of the BRST transformations to solutions of
the field equations.

Starting from the Lagrangian density, one way to find the BRST
generator is to compute the Noether current $J^a$. Using left BRST
transformations $s^L$ ($s^Lz=(-)s^Rz$ for even (odd) fields $z$),
\begin{equation}
J^a=s^L z{\partial L\over \partial z_{;a}}-j^a,
\end{equation}
where $s^LL=j_a^{\ ;a}$.
For the present example,
\begin{equation}
s^LA_a=c_{;a},\quad s^L\overline c=b,\quad s^Lc=s^Lb=0,\quad
j_a=-b\,c_{;a}
\end{equation}
The Noether current is therefore
\begin{equation}
J^a=-bc^{;b}+F^{ab}c_{;b}\label{nc}
\end{equation}
The Noether charge $\Omega$ is the volume integral of the local charge
density $\omega=n_aJ^a$.

Decomposition of the Lagrangian density, following the outline given in
the appendix, results in the momenta,
\begin{eqnarray}
\pi^a&=&g^{ab}(\phi_{|b}-{\cal L}\phi_a),\quad\pi=-b\nonumber\\
p&=&{\cal L}\overline c,\quad\overline p=-{\cal L}c.
\end{eqnarray}
Using these expressions, equation(\ref{nc}) leads trivially to the BRST
charge density
\begin{equation}
\omega=\overline p b-c\pi^a_{\,|a},\label{omg}
\end{equation}
In the notation used in the previous section, $\overline E=-\pi$
(Lorentz gauge condition) and $E=\pi^a_{\,|a}$ (Gauss' law
constraint).

The vanishing-ghost boundary condition is given by
\begin{equation}
c=\overline c=b=\pi=0,\quad \phi_a=0.
\end{equation}
Using equation (\ref{aux}) and eliminating momenta puts this into mixed
form,
\begin{eqnarray}
c=\overline c=0\nonumber\\
{\cal L}\phi+K\phi=0\nonumber\\
\phi_a=0
\end{eqnarray}
This set of boundary conditions fixes the magnetic field on the
boundary.

No other linear combination of $\overline E$ and $E$ can be put into
mixed form, except for $E$ itself, which is a total divergence. The
momentum $\pi$ does not appear in $E$ and $\phi$ can be fixed on the
boundary by rule 3 of section 2. The only other set of mixed boundary
conditions is therefore
\begin{eqnarray}
{\cal L} c={\cal L} {\overline c}=0\nonumber\\
{\cal L}\phi_a=0\nonumber\\
\phi=0
\end{eqnarray}
This set of boundary conditions fixes the electric field on the
boundary.

\section{Linearised Gravity}

Linearised gravity forms the starting point for order $\hbar$ quantum
gravity calculations based on Einstein gravity, as well as having wider
applications to supergravity and superstring theories by taking various
spacetime dimensions. We are seeking sets of local boundary conditions
for the path integral, using t'Hooft-Veltman gauges because they are
widely used and covariant.

For t'Hooft-Veltman gauges, the metric fluctuation $\gamma_{ab}$ is
accompanied by ghost fields $C_a$ and antighost fields $\overline C^a$.
The metric fluctuation is defined in terms of the perturbed metric
\begin{equation}
g_{ab}+2\kappa\gamma_{ab},
\end{equation}
where $\kappa^2=8\pi G$. We will also make use of the dual quantity
\begin{equation}
\overline\gamma^{ab}=g^{(ab)(ef)}\gamma_{ef},
\end{equation}
defined by the metric
\begin{equation}
g^{(ab)(cd)}=\case1/2(g^{ac}g^{bd}+g^{ad}g^{bc}-g^{ab}g^{cd}).
\end{equation}

In order to set up a phase space associated with the hypersurface
$\Sigma$ we need to decompose all of these fields into normal and
tangential components,
\begin{eqnarray}
\gamma_{ab}&=&\phi_{ab}+2\phi_{(a} n_{b)}+\phi n_an_b\nonumber\\
\overline\gamma_{ab}&=&\overline\phi_{ab}+2\overline\phi_{(a}
n_{b)}+\overline\phi n_an_b\nonumber\\
C_a&=&c_a+c n_a,\quad \overline C^a=\overline c^a+\overline c n^a
\end{eqnarray}
(The index structure distinguishes different vector and scalar
quantities.) Momenta conjugate to $\phi_X$, $c_X$ and $\overline c^X$
are denoted by $\pi^X$, $p^X$ and $\overline p_X$ respectively.

The background metric on $\Sigma$ will be denoted by $h_{ab}$.
Variations in the surface metric correspond to variations in both
$\phi_{ab}$ and $\phi_a$, but variations in the surface geometry depend
only on $\phi_{ab}$.

The Lagrangian density $L$ can be taken to be the sum of two terms, the
gauge-fixed Einstein-Hilbert term and the ghost terms. For a
t'Hooft-Veltman gauge-fixing term,
\begin{eqnarray}
L_\gamma&=&-\case1/2\overline\gamma^{ab;c}\gamma_{ab;c}
+R^{acbd}\,\overline\gamma_{ab}\gamma_{cd}
+G^{ac}g^{bd}\,\overline\gamma_{ab}\gamma_{cd}\nonumber\\
L_{gh}&=&-\overline C^{a;b}C_{a;b}
+R_a^{\ b}\,\overline C^a C_b\label{grl}
\end{eqnarray}
where $G_{ab}$ is the Einstein tensor. The auxilliary field $b$ has
already been eliminated.

The non-vanishing BRST transformations are
\begin{equation}
s^L\gamma_{ab}=2C_{(a;b)},\quad s^L\overline
C=2\overline\gamma^{ab}_{\ \ ;b}.
\end{equation}
The BRST charge density can be calculated as in the last section, using
equation (\ref{q}) and the decompositions in Appendix A. The result can
be written in the form
\begin{equation}
\omega=\overline p^a\,\overline E_a+\overline p\,\overline
F+c_a\,E^a+c\,F.
\end{equation}
Explicit expressions for $\overline E_a$, $\overline F$, $E^a$ and $F$
appear in appendix C, equations (\ref{ebargrav})-(\ref{fgrav}). Whilst
$\overline E_a$ and $\overline F$ are already in the correct form
(given in equation \ref{rn}), $E^a$ and $F$ are not. (Even in this form
they can be used to obtain non-local boundary conditions which are
potentially useful for particular backgrounds.)

We still have the freedom to perform the linear transformations
described in rule 1 at the end of section 2. We first of all perform
linear transformations on $E^a$ and $F$ to separate divergences of
momenta from gradients of momenta,
\begin{eqnarray}
F'&=&F-\overline E_c^{\ |c}+\alpha K\overline F\\
E_a'&=&E_a-K_a^{\ c}\overline E_c-\beta\overline F_{|a}.
\end{eqnarray}
The new $F'$ commutes with $\phi_a$ and the fields
\begin{eqnarray}
\phi^K_{ab}&=&\phi_{ab}-K^{-1}K_{ab}\phi_c^{\ c}\nonumber\\
\phi^{(\alpha)}&=&\phi+\alpha\phi_a^{\ a}.
\end{eqnarray}
A final linear transformation allows a choice of $\overline F'$ and
$\overline E_a'$ from the set $\{F',E_a',\overline F,\overline E_a\}$.

What happens depends very much on whether the extrinsic curvature
$K_{ab}$ is proportional to the surface metric. If $K_{ab}=Kh_{ab}/3$,
then we have the following boundary conditions:
\begin{itemize}
\item[I] $\{\overline E_a,\overline F,\phi_{ab},\overline c^a,\overline
c,c_a,c\}=0$
\item[II] $\{\overline E_a,F',\phi^K_{ab},\phi^{(\alpha)},c,\overline
c_a,p+\alpha K\overline c,\overline p-\alpha Kc\}=0$
\end{itemize}
If, in addition, $K$ is constant then
\begin{itemize}
\item[III] $\{E'_a,\overline F,\phi_a,\overline
c,c,p_a-K_a^{\ b}\overline c_b,
\overline p_a+K_a^{\ b} c_b\}=0$
\item[IV] $\{E'_a,F',\phi_a,p+\alpha K\overline c,\overline p-\alpha
Kc,
p_a-K_a^{\ b}\overline c_b-\beta\overline c_{|a},\overline p_a+
K_a^{\ b} c_b+\beta c_{|a}\}=0$
\end{itemize}
In cases (III) and (IV), the expression for $E'_a$ is a total
divergence which can be integrated to obtain boundary conditions of the
correct type.

The boundary conditions are written explicitly in table \ref{taba}.
Boundary conditions (I) have been applied previously to applications in
quantum cosmology. The other boundary conditions are new, to the best
of our knowledge. Boundary conditions (III) are especially interesting
because they contain no spatial derivative terms.

Difficulties arise when the extrinsic curvature is not proportional to
the surface metric. The function $E'_a$ can be written as a total
divergence, but not of a symmetric tensor (see equation(\ref{epr})).
Boundary conditions (III) and (IV) belong to a wider class of boundary
conditions where the projection operators in equations (\ref{rn}) and
(\ref{dt}) include surface derivatives. This leaves boundary conditions
(I) and (II). The resulting expressions are listed in table
\ref{tabb}.

\section{CONCLUSIONS}

We have assumed that the boundary conditions on the path integral are
local, linear and BRST invariant. Locality means that the boundary
conditions at a point depend only on the fields and their derivatives
and has been imposed because it is useful for quantum field theory with
non-trivial background fields. Linearity is imposed for the same
reason, since linear theory is the starting point of the
$\hbar$-expansion in quantum field thoery.

BRST invariance is meant in the sense that the BRST operator anihilates
the result of the path integral. The boundary conditions themselves are
BRST invariant in the sense that, when written in terms of momenta,
they commute with the BRST generator.

With these assumptions, the boundary conditions can all be generated by
following the rules given at the end of section 2. Using these rules it
has been possible to find all of the boundary conditions for linearised
gravity with t'hooft-Veltmann gauge fixing that are of the mixed
Robin-Dirichlet type, generalised to include surface derivative terms.
These are given in tables I and II. Set I of boundary conditions which
fix the surface geometry is known already \cite{b3,e4,e5,e6} and the
other sets are new. Set III has no surface derivatives.

Boundary conditions for linearised gravity are useful in quantum
cosmology. The first set of boundary conditions fix the scale factor of
the universe. The second set of boundary conditions would correspond to
fixing the expansion rate of the universe instead of the scale-factor.
The expansion rate has the advantage over the scale-factor of being a
single-valued function of time in classical cosmological models.

\appendix
\section{HYPERSURFACES}

Introducing a hypersurface $\Sigma$ into the manifold ${\cal M}$ leads
to a natural decomposition of the tangent space of ${\cal M}$ into the
tangent space of $\Sigma$ and its compliment along the normal vector
$n^a$. We denote the intrinsic metric by
\begin{equation}
h_{ab}=g_{ab}-n_an_b.
\end{equation}
The Lie derivative of the intrinsic metric along the normal direction
defines the extrinsic curvature $K_{ab}$,
\begin{equation}
{\cal L}h_{ab}=2K_{ab}
\end{equation}
The covariant derivative on ${\cal M}$, expressed by $\phi_{a;b}$,
induces a covariant derivative on $\Sigma$. The definition
\begin{equation}
\phi_{a|b}=\phi_{a;b}-n_b{\cal
L}\phi_a+\Gamma^c_{\ ab}\phi_c,\label{decomp}
\end{equation}
where
\begin{equation}
\Gamma^c_{\ ab}=K^c_{\ a}n_b+K^c_{\ b}n_a+({\cal L}n)^cn_an_b.
\end{equation}
is particularly useful. This expression extends to tensors on $\Sigma$.
A particular example is the surface metric itself, which is easily seen
to satisfy $h_{ab|c}=0$.

Decomposition of the Riemann tensor is straight-forward if we take
${\cal L}n=0$. Two applications of equation (\ref{decomp}) gives
\begin{eqnarray}
R_{abc0}&=&K_{cb|a}-K_{ca|b}\nonumber\\
R_{a0b0}&=&K_a^{\ c}K_{cb}-{\cal L}K_{ab}\nonumber\\
R_{abcd}&=&r_{abcd}-K_{ac}K_{bd}+K_{ad}K_{bc}
\end{eqnarray}
where $r^a_{\ bcd}$ is the Riemann tensor for $h_{ab}$.

\section{MOMENTA}

Equation (\ref{decomp}) can be used to express any Lagrangian that is
second order in derivatives as a function $L({\cal L}\phi_X,\phi_X)$.
Momenta $\pi^X$ are defined by differentiation of Lagrangian densities
$L$ with respect to ${\cal L}\phi_X$,
\begin{equation}
\pi^X={1\over \det g}{\partial (L \det g)\over \partial ({\cal
L}\phi_X)}.\label{momda}
\end{equation}
Because of the linear form of equation (\ref{decomp}), it is also
possible to write this as
\begin{equation}
\pi^X=n_a{\partial L\over \partial \phi_{X;a}}.\label{momdb}
\end{equation}

For gauge-fixed Electrodynamics in curved spaces with the Lagrangian
given by equations (\ref{edl}),
\begin{eqnarray}
L_A&=&-\case1/2g^{ab}(\phi_{|a}-{\cal L}\phi_a)
(\phi_{|b}-{\cal L}\phi_b)+\dots\nonumber\\
L_{gh}&=&-({\cal L}\overline c)({\cal L}c)+\dots\nonumber\\
L_{gf}&=&-b(\phi_a^{\ |a}+{\cal L}\phi+K\phi)+\case1/2b^2.
\end{eqnarray}
These allow the momenta to be read off using equation (\ref{momda}).

For gravity with the Lagrangian density given by equations (\ref{grl})
it is best to use equation (\ref{momdb}),
\begin{equation}
\pi^X=-n^c\overline \gamma^{ab}_{\ \ ;c},\quad
p^X=n^b\overline C^a_{\ ;b},\quad\overline p_X=n^b C_{a;b}
\end{equation}
After application of equation (\ref{decomp}), the momenta become
\begin{eqnarray}
\pi_{ab}&=&-({\cal
L}\overline\phi_{ab}-2K_{(a}^{\ c}\overline\phi_{b)c})\\
\pi^a&=&-2g^{ab}({\cal L}\phi_b-K_b^{\ c}\phi_b)\\
\pi&=&-{\cal L}\overline\phi\\
p^a&=&+({\cal L}\overline c^a+K^a_{\ b}\overline c^b),
\quad p={\cal L}\overline c\\
\overline p_a&=&-({\cal L} c_a-K_a^{\ b} c_b),\quad
\overline p=-{\cal L}c.\label{momg}
\end{eqnarray}

\section{BRST CHARGE FOR GRAVITY}

Under the BRST variations, the Lagrange densities (\ref{grl}) transform
by a divergence plus extra terms,
\begin{equation}
s^LL=j^a_{\ ;a}+
2E^{ab}\left(2C^d_{\ ;b}\gamma_{ad}+C^d\gamma_{ab;d}\right),
\end{equation}
where
\begin{equation}
j_a=-2\overline\gamma_{ab;c}C^{b;c}+
2(R_{a\ d}^{\ b\ c}+R_d^{\ b}\delta^c_{\ a}-
E^{bc}g_{ad})\overline\gamma_{bc}C^d
\end{equation}
The tensor $E^{ab}$ depends on the Einstein tensor of the background
fields and also the stress-energy tensor if a matter Lagrangian is
included,
\begin{equation}
E^{ab}=G^{ab}-\kappa^2 T^{ab}.
\end{equation}
This tensor vanishes for background fields that satisfy the Einstein
equations, which will be assumed throughout.

The BRST generator $\omega$ can be obtained from the Noether current,
$2\omega=n^cJ_c$, where
\begin{equation}
J^c={\partial L\over\partial \gamma_{ab;c}}s^L\gamma_{ab}+
s^L\overline C^a{\partial L\over\partial \overline C^a_{\ ;c}}-j^c
\end{equation}
For the Lagrange densities (\ref{grl}), this becomes
\begin{equation}
J_c=-2\overline\gamma_{ab;c}C^{a;b}-2\overline\gamma^{ab}_{\ \ |b}C_{a;c}-j_c.
\end{equation}
Using the decomposition rule (\ref{decomp}) and the momenta
(\ref{momg}), the BRST generator can be written in the form
\begin{equation}
\omega=\overline p^a\,\overline E_a+\overline p\,\overline
F+c_a\,E^a+c\,F.
\end{equation}
The functions appearing here are evaluated on phase space
$(\pi^X,\phi_X)$. The dependence of the functions on the momenta is
given explicitly by
\begin{eqnarray}
\overline E_a(\pi^X,0)&=&-\case1/2\pi^a\label{ebargrav}\\
\overline F(\pi^X,0)&=&-\pi\\
E_a(\pi^X,0)&=&-\pi_{ab}^{\ \ |b}-\case1/2K_{ab}\pi^b\\
F(\pi^X,0)&=&K_{ab}\pi^{ab}-\case1/2\pi^a_{\ |a}\label{fgrav}
\end{eqnarray}
For the boundary conditions we need to eliminate the momenta. This
leads to the following expressions:
\begin{eqnarray}
\overline E_a&=&{\cal
L}\phi_a+K\phi_a+\nabla^b\overline\phi_{ab}\label{ebarb}\\
\overline F&=&{\cal
L}\overline\phi+K\overline\phi-K^{ab}\overline\phi_{ab}+
\nabla^a\phi_a
\end{eqnarray}
The linear combinations of $E_a$ and $F$ that come closest to the form
that we require are
\begin{eqnarray}
E_a-K_a^{\ b}\overline E_b&=&
\nabla^b{\cal L}\overline\phi_{ab}+
2\nabla^b(K_a^{\ c}\overline\phi_{bc}-\overline\phi h_{ab})
+K^{bc}_{\ \ |a}(\overline\phi_{bc}+\overline\phi h_{bc})\nonumber\\
&&+(K_a^{\ c}K_c^{\ b}-r_a^{\ b}-\nabla^2)\phi_b\label{epr}\\
F-\overline E_a^{\ |a}&=&
-K^{ab}{\cal L}\overline\phi_{ab}+({\cal
L}K_{ab}-\nabla_a\nabla_b)(\overline\phi^{ab}+\overline\phi
h^{ab})\nonumber\\
&&+(K_b^{\ a|b}-K^{|a}+2K^{ab}-2Kh^{ab}+K\nabla^a)\phi_a\label{fpr}
\end{eqnarray}
Surface derivatives on $\phi_X$ are denoted now by $\nabla^a$.

\begin{table}
\begin{tabular}{llll}
Set&$n$&${\cal Q}$&${\cal R}$\\
\hline\hline
I&1&$\phi_{ab}$&${\cal L}\phi_a+K\phi_a-\case1/2\nabla_a\phi$\\
&&$c_a,c$&${\cal L}\overline\phi+2K\overline\phi+\nabla^a\phi_a$
\hfill$=\overline F$\\
\hline
II&2&$\phi+\alpha \phi^T$,$\phi^K_{ab}$&${\cal
L}\phi_a+K\phi_a+\nabla_a(\phi^L-\overline\phi)$\\
&&&$K{\cal L}\phi^L-\Delta\phi^L-(K_L^{|a}+
2K_L\nabla^a)\phi_a-(\alpha+1)K\overline F$\\
&&$c_a$&${\cal L} c+\alpha Kc$\\
\hline
III&0&$\phi+\beta \phi^T$,$\phi_a$&${\cal
L}\phi^L_{ab}-K_L\phi^L_{ab}$\\
&&&${\cal L}{\overline \phi}+2K\overline\phi-K\phi^L$\\
&&$c$&${\cal L} c_a+\beta Kc_a$\\
\hline
IV&2&$\phi_a$&$K{\cal
L}\phi^L-\Delta\phi^L-(K_L^{|a}+2K_L\nabla^a)\phi_a-(\alpha+1)K\overline
F$\\
&&&${\cal L}\phi^L_{ab}-K_L\phi^L_{ab}-(\beta+1)\overline F$\\
&&&${\cal L} c_a+\alpha K c_a+\beta K\nabla_a c$\\
\end{tabular}
\caption{Four sets of boundary conditions for linearised gravity with
extrinsic curvature $K_{ab}=Kh_{ab}/3$. Each entry is equated to zero,
quantities listed under ${\cal Q}$ denoting dirichlet boundary
conditions which are combined with the entries under ${\cal R}$ to form
the mixed class ${\cal M}_n$. Special combinations of fields are
denoted by $\phi^T=h^{ab}\phi_{ab}$,
$\phi^K_{ab}=\phi_{ab}-(1/K)\phi^TK_{ab}$,
$\phi^L_{ab}=\phi_{ab}-\phi^Th_{ab}$ and $\phi^L=-(2/3)\phi^T$. The
operator $\Delta=({\cal L}K)+K^2-\nabla^2$ and $\overline F$ is defined
in the second line.}
\label{taba}
\end{table}

\begin{table}
\begin{tabular}{llll}
Set&$n$&${\cal Q}$&${\cal R}$\\
\hline\hline
I&1&$\phi_{ab}$&${\cal L}\phi_a+K\phi_a-\case1/2\nabla_a\phi$\\
&&$c_a,c$&${\cal L}\overline\phi+2K\overline\phi+\nabla^a\phi_a$
\hfill$=\overline F$\\
\hline
II&2&$\phi+\alpha \phi^T,\phi^K_{ab}$&${\cal
L}\phi_a+K\phi_a+\nabla^b\overline\phi_{ab}$\\
&&&$K{\cal L}\phi^L-\Delta^{ab}\phi^L_{ab}-
2K^L_{ab}K^{ab}\overline\phi-((K_L)^{ab}_{\ \ |b}+2(K_L)^{ab}\nabla_b)\phi_a-(\alpha+1)K\overline
F$\\
&&$c_a$&${\cal L} c+\alpha Kc$\\
\end{tabular}
\caption{Two sets of boundary conditions for linearised gravity with
extrinsic curvature $K_{ab}\ne Kh_{ab}/3$. Each entry is equated to
zero as before. Special combinations of fields are as in table 1,
except for $\phi^L=K^{-1}K^{ab}\phi^L_{ab}$ and $\Delta^{ab}=(2{\cal
L}K^{ab}-K^{-1}({\cal
L}K)K^{ab}-4K^{ac}K_c^{\ b}+KK^{ab}-\nabla^a\nabla^b)$.}
\label{tabb}
\end{table}

\acknowledgments
P. Silva is supported by the Government of Venezuela.

\end{document}